\documentclass[10pt,letterpaper]{article}
\usepackage[top=0.85in,left=2.75in,footskip=0.75in,marginparwidth=2in]{geometry}

\usepackage[utf8]{inputenc}

\usepackage{cite}

\usepackage{nameref,hyperref}

\usepackage[right]{lineno}

\usepackage{microtype}
\DisableLigatures[f]{encoding = *, family = * }

\raggedright
\usepackage{changepage}

\usepackage[aboveskip=1pt,labelfont=bf,labelsep=period,singlelinecheck=off]{caption}

\makeatletter
\renewcommand{\@biblabel}[1]{\quad#1.}
\makeatother

\usepackage{lastpage,fancyhdr,graphicx}
\usepackage{epstopdf}
\usepackage{amsfonts}
\usepackage{amsmath}

\fancyheadoffset[L]{2.25in}
\fancyfootoffset[L]{2.25in}

\usepackage{color}

\definecolor{Gray}{gray}{.25}

\usepackage{graphicx}

\usepackage{sidecap}

\usepackage{wrapfig}
\usepackage[pscoord]{eso-pic}
\usepackage[fulladjust]{marginnote}
\reversemarginpar

\begin{document}
	\vspace*{0.35in}
	
	\begin{flushleft}
		{\Large
			\textbf\newline{Superoscillatory $\mathcal{PT}$-symmetric potentials}
		}
		\newline
		\\
		Yaniv Eliezer\textsuperscript{1,$\dagger$},
		Alon Bahabad\textsuperscript{1},
		Boris A. Malomed\textsuperscript{1}
		\\
		\bigskip
		\bf{1} Department of Physical Electronics, School of Electrical
		Engineering, Fleischman Faculty of Engineering,\\ and Center of
		Light-Matter Interaction, \\ Tel-Aviv University, Tel-Aviv 69978, Israel \\
		\bigskip
		$\dagger$ yaniveli@post.tau.ac.il
		
	\end{flushleft}
	
	\section*{Abstract}
	We introduce the one-dimensional $\mathcal{PT}$-symmetric Schr\"{o}dinger
	equation, with complex potentials in the form of the canonical
	superoscillatory and suboscillatory functions known in quantum mechanics and
	optics. While the suboscillatory-like potential always generates an entirely
	real eigenvalue spectrum, its counterpart based on the superoscillatory wave
	function gives rise to an intricate pattern of $\mathcal{PT}$%
	-symmetry-breaking transitions, controlled by the parameters of the
	superoscillatory function. One scenario of the transitions proceeds smoothly
	via a set of threshold values, while another one exhibits a sudden jump to
	the broken $\mathcal{PT}$ symmetry. Another noteworthy finding is the
	possibility of restoration of the $\mathcal{PT}$ symmetry, following its
	original loss, in the course of the variation of the parameters.
	
	
	\section{\label{sec:level0} Introduction}
	
	The concept of complex-valued quantum Hamiltonians has been known for
	decades \cite%
	{feshbach1954model,lindblad1976generators,PhysRevD.18.2914,doi:10.1080/00268977800102631}%
	, yet for a long time it was commonly believed that a mandatory requirement
	for the reality of eigenvalues in a quantum system was the Hermiticity of
	the Hamiltonian. This belief was held firm in spite of producing several
	examples showing that complex Hamiltonians may generate a set of real
	eigenvalues \cite%
	{caliceti1980perturbation,andrianov1982large,hollowood1992solitons,scholtz1992quasi}%
	. It was the seminal work of Carl Bender and Stefan Boettcher (1998) \cite%
	{bender1998real} which showed that, by replacing the Hermiticity with the
	weaker condition of the $\mathcal{PT}$ symmetry, it is possible to construct
	classes of non-Hermitian Hamiltonians that exhibit completely real spectra
	of eigenvalues. An important principle found in this context is the
	necessary, yet not sufficient, condition for the $\mathcal{PT}$ symmetry,
	which states that a complex potential, if it is a part of the Hamiltonian,
	must satisfy the constraint $V(\xi )=V^{\ast }\left( {-\xi }\right) $, where $%
	\xi $ is the position coordinate. Many works examined different families of
	complex potentials satisfying this condition \cite%
	{bender1998real,bender1999complex,znojil2001pt,bender2002complex,khare2004analytically,khare2006complex,moiseyev2011non,bender2016rigorous}.
	
	In the field of optics, the paraxial propagation of light in materials which
	include optical gain and loss can be modelled by the Schr{\"{o}}dinger
	equation including a complex potential. As a consequence, it is possible to
	emulate the evolution of quantum $\mathcal{PT}$-symmetric systems in terms
	of classical optics. This concept was elaborated in numerous works \cite
	{berry1998diffraction, ruschhaupt2005physical, el2007theory,
		berry2008optical,
		klaiman2008visualization,longhi2009bloch,kottos2010optical,driben2011stability,ramezani2012exceptional,suchkov2012wave,alexeeva2012optical,PhysRevA.82.043803,sukhorukov2010nonlinear,longhi2012photonic,longhi2014optical,chong2011p,pilozzi2017topological} 
	and demonstrated experimentally in various settings, such as optical
	waveguides \cite%
	{PhysRevLett.103.093902,ruter2010observation,regensburger2012parity}, lasing
	\cite{hodaei2014parity}, microcavity resonators \cite{peng2014parity},
	metamaterials \cite{feng2013experimental}, microwaves \cite{micro},
	electronic circuits \cite{schindler2011experimental} and acoustics \cite%
	{fleury2015invisible}.{\LARGE \ }
	
	Usually, $\mathcal{PT}$-symmetric potentials contain a control parameter,
	the variation of which leads to breaking of $\mathcal{PT}$ symmetry, at
	a certain threshold value of the parameter. Above the threshold, eigenstates
	of the $\mathcal{PT}$-symmetric Hamiltonian no longer remain eigenfunctions
	of the $\mathcal{PT}$ operator, and, at least, a subset of the spectrum of
	eigenvalues ceases to be real \cite{bender1998real,bender2007making}. $%
	\mathcal{PT}$-symmetry breaking was theoretically considered in various
	contexts, and experimentally realized in optics \cite{makris2008beam,PhysRevA.81.063807,makris2011mathcal,PhysRevLett.103.093902}.
	
	Superoscillations are a phenomenon in which a band-limited signal oscillates
	locally faster than its highest Fourier component \cite{berry1994faster}. A
	canonical superoscillatory function was found by Aharonov \textit{et al}.
	\cite{aharonov1988result} in the theoretical framework of weak quantum
	measurements. A complementary canonical form for\textit{\ suboscillatory}
	functions, i.e., signals which exhibit local oscillations that are slower
	than their lowest Fourier component, was recently found as well \cite
	{suboscllation}. Superoscillations have found applications in various fields
	of optics, such as imaging \cite{huang2007optical,rogers2012super,wong2013optical},
	ultrafast optics \cite{eliezer2017breaking,eliezer2018experimental},
	nonlinear light propagation \cite{remez2015super}, light-beam shaping \cite%
	{greenfield2013experimental,eliezer2016super,zacharias2017axial} and optical
	traps \cite{singh2017particle}.
	
	In this work we first derive a potential for which the canonical
	superoscillatory function is an eigenstate of the Hamiltonian of the Schr{%
		\"{o}}dinger equation. The potential turns out to be a canonical
	suboscillatory function, and it is endlessly $\mathcal{PT}$-symmetric,
	always generating an entirely real spectrum of eigenvalues. Then we consider
	the superoscillatory complex canonical function itself as a new complex $%
	\mathcal{PT}$-symmetric potential. Varying its parameters, we report an
	intricate picture of $\mathcal{PT}$-symmetry-breaking phase transitions. In
	particular, there are regions in the parameter space where the variation
	leads to gradual expansion of the complex ($\mathcal{PT}$-symmetry-broken)
	part of the spectrum, while in other regions one can find paths for changing
	the parameters that lead to abrupt $\mathcal{PT}$-symmetry breaking.
	Furthermore, there are regions in which the initial symmetry breaking is
	followed by its restoration. The effects of broken $\mathcal{PT}$ symmetry
	on the evolution of localized field pulses are explored too by means of
	direct simulations.
	
	
	\section{\label{sec:level1} Analysis}
	
	\subsection{A complex potential supporting the superoscillating wave function%
	}
	
	First we consider the canonical superoscillatory function, devised
	originally in the context of weak measurements in quantum mechanics \cite%
	{aharonov1988result}\cite{berry2006evolution}:
	\begin{equation}
	f_{\mathrm{SO}}(\xi )=\left[ \cos (\xi )+ia\sin (\xi )\right] ^{N}\equiv %
	\left[ g(\xi )\right] ^{N},  \label{eqso1}
	\end{equation}%
	where $a$ is a real parameter, and $N$ is an integer. We aim to identify it
	as a stationary wave function,
	\begin{equation}
	\psi (\xi ,\eta )=e^{iE\eta }\left[ \cos (\xi )+ia\sin (\xi )\right] ^{N},
	\label{eq1}
	\end{equation}%
	of the scaled Schr{\"{o}}dinger equation with a complex potential, $V(\xi )$%
	:
	\begin{equation}
	i\psi _{\eta }=\psi _{\xi \xi }+V(\xi )\psi .  \label{schrodinger}
	\end{equation}%
	In terms of optics realization, $\eta $ and $\xi $ are, respectively,
	the longitudinal propagation distance and transverse spatial coordinate,
	while $E$ is the propagation constant \cite{yariv1989quantum} ($-E$ would be
	the energy eigenvalue in the quantum model).
	
	By substituting expression (\ref{eq1}) into Eq. (\ref{schrodinger}), we
	conclude that the wavefunction (\ref{eq1}) is supported as an eigenstate,
	with the eigenvalue $E=N^{2}$, by the following potential:
	\begin{equation}
	V_{\mathrm{SO}}(\xi )=\frac{(a^{2}-1)N(N-1)}{\left[ \cos (\xi )+ia\sin (\xi )%
		\right] ^{2}}.  \label{eq2}
	\end{equation}
	
	The complex potential $V_{\mathrm{SO}}(\xi )$ is a $\mathcal{PT}$-symmetric one,
	as it is subject to the condition $V_{\mathrm{SO}}(\xi )=V_{\mathrm{SO}%
	}^{\ast }(-\xi )$, with $\ast $ standing for complex conjugation \cite%
	{bender2007making}. In addition, this potential, which supports the
	canonical superoscillatory function as the stationary state of Schr\"{o}%
	dinger equation (\ref{schrodinger}), can be identified as the known
	canonical suboscillatory function \cite{suboscllation}.
	
	Because $V_{\mathrm{SO}}(\xi )$ is a complex periodic function, it can be
	expanded into the Fourier series:
	\begin{equation}
	V_{\mathrm{SO}}(\xi )=(a^{2}-1)N(N-1)\sum_{m=-\infty }^{+\infty }C_{m}\exp
	(im\xi ),  \label{vsoperiodic}
	\end{equation}%
	where the coefficients $C_{n}$ can be readily calculated:
	\begin{equation}
	\begin{split}
	& C_{m}=\frac{1}{2\pi }\int_{-\pi }^{+\pi }\frac{\exp (-im\xi )}{\left[ \cos
		(\xi )+ia\sin (\xi )\right] ^{2}}d\xi \\
	& =\left\{ {%
		\begin{array}{ccccccccccccccccccc}
		{\ }-2m\left( a+1\right) ^{\frac{m}{2}-1}\left( a-1\right) ^{-\left( \frac{m%
			}{2}+1\right) }, & {m\in \{\mathrm{even}{<0\},}}  \\
		0, & \mathrm{otherwise.}
		\end{array}%
	}\right.
	\end{split}
	\label{cn1}
	\end{equation}%
	As shown by Bender \textit{et al}. \cite{bender1999complex}, any complex
	potential having a polynomial form in $\exp (i\xi )$ results in an entirely
	real spectrum. According to Eq. (\ref{cn1}), the Fourier transform of $V_{%
		\mathrm{SO}}(\xi )$ is discrete and completely single-sided, i.e., the
	potential (\ref{vsoperiodic}) is a polynomial of this type, hence, for all $%
	a $ and $N$, the respective energy spectrum remains endlessly real, with no
	occurrence of $\mathcal{PT}$-symmetry breaking. An example of potential $V_{%
		\mathrm{SO}}(\xi )$ with parameters $N=4$ and $a=2$ is shown in Fig. \ref%
	{fig:vsoEbands}. Note that the calculated energy bands for this case are
	indeed entirely real, and they include eigenvalue $E=N^{2}=16$,
	corresponding to the above-mentioned eigenfunction in the form of the
	canonical superoscillatory function.
	
	\begin{figure}[ht] 
		\includegraphics[width=0.7\linewidth,trim={0.7cm 0.0cm 1.3cm
			0.0cm},clip]{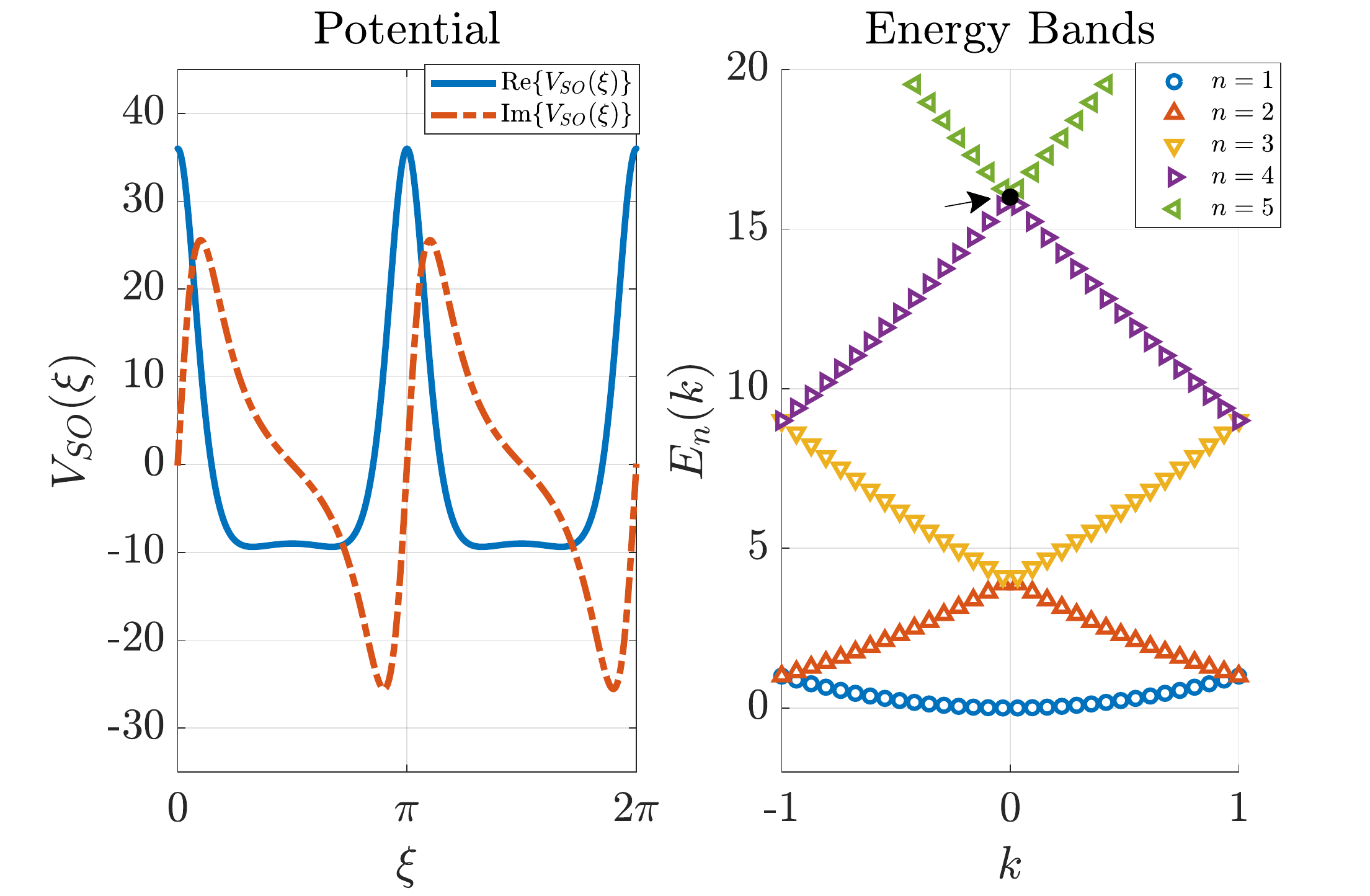}
		\caption{ \textbf{Energy bands structure (right) of potential $\mathbf{%
					V_{SO}(\protect\xi )}$ (left), calculated for $\mathbf{N=4}$, $\mathbf{a=2}$. 
				A black arrow marks the location where $E=N^{2}$. }}
	\label{fig:vsoEbands}
	\end{figure}	
	
	\newpage 
	
	\subsection{The complex potential in the form of the canonical
		superoscillating function}
	
	We now consider the complex potential which itself has the form of the
	canonical superoscillatory function,
	\begin{equation}
	V_{\mathrm{SOF}}(\xi )=\left[ \cos (2\xi )+ia\sin (2\xi )\right] ^{N},
	\label{vsof}
	\end{equation}%
	where the above constraints on the parameters $a$ and $N$ are relaxed, both
	being taken as a pair of positive rational numbers. This form can be
	regarded as a generalization of previously examined real and $\mathcal{PT}$%
	-symmetric potentials, \textit{viz}., ones in the form of $\cos ^{N}(\xi )$
	(for $a=0$ and positive integer $N$), $i\sin ^{N}(\xi )$ (for $a\gg 1$ and
	odd integer $N$) \cite{bender1999complex,bender2007making}, $4\left[ \cos
	^{2}(\xi )+ia\sin (2\xi )\right] $ ($a>1,N=1$) \cite{PhysRevA.81.063807} and
	$\exp (iN\xi )$ (for $a=1$ and positive integer $N$) \cite%
	{cannata1998schrodinger,bender1999complex}. Since $N$ may now be a
	fractional power, the complex potential (\ref{vsof}) may be a multivalued
	function. To remove the complex-roots ambiguity, we select the following
	relation to uniquely define $V_{\mathrm{SOF}}(\xi )$: {\footnotesize
		\begin{equation}
		\begin{split}
		& {\displaystyle V_{\mathrm{SOF}}\left( \xi \right) =\exp \left( {N\left[ {%
					\ln |g\left( \xi \right) |+i\,\mathrm{atan}_{2}\left( {\frac{{{\mathop{\rm
										Im}\nolimits}\left\{ {g\left( \xi \right) }\right\} }}{{{\mathop{\rm Re}%
									\nolimits}\left\{ {g\left( \xi \right) }\right\} }}}\right) }\right] }%
			\right) } \\
		& =\exp \left( {N\left\{ {\frac{1}{2}\ln \left[ {{{\cos }^{2}}\left( {2\xi }%
					\right) +{a^{2}}{{\sin }^{2}}\left( {2\xi }\right) }\right] +i\,\mathrm{atan}%
				_{2}\left( {a\tan \left( {2\xi }\right) }\right) }\right\} }\right) ,
		\end{split}%
		\end{equation}%
	}%
	where $g(\xi )$ is the same as in Eq. (\ref{eqso1}), and $atan_{2}(\cdot )$
	is the \textquotedblleft four-quadrant" inverse tangent, which is defined as
	the angle between the positive $x$ axis and the vector ending at point $%
	\left( \mathrm{Re}\{g(\xi )\},\mathrm{Im}\{g(\xi )\}\right) $. The resulting
	angle belongs to the interval [0, $\pi $] for $\mathrm{Im}\{g(\xi )\}\geq 0$,
	and to ($-\pi $, 0) for $\mathrm{Im}\{g(\xi )\}<0$.
	
	We use the numerically implemented Bloch-Floquet technique \cite%
	{ashcroft2011solid} to calculate the energy spectrum of potential (\ref{vsof}%
	) for various values of the parameters $a$ and $N$. Unlike the complex potential
	of Eq. (\ref{eq2}), the present one gives rise to $\mathcal{PT}$-symmetry breaking at
	sufficiently large values of $a$. To quantify this effect, we introduce a
	measure, $\rho _{n}(a,N)$, for the $n$-th band of eigenvalues $E$, which
	quantifies a relative degree of the $\mathcal{PT}$-symmetry breaking in the band, by
	calculating the portion of the Brillouin zone, $-1<k<+1$ (of the
	corresponding quasi-momentum $k$) in which the eigenvalues are complex:%
	{\normalsize {\LARGE \ }
		\begin{equation}
		\rho _{n}(a,N)=\frac{1}{2}\int\limits_{-1}^{+1}I_{n}\left( k,a,N\right) dk,
		\label{ratio}
		\end{equation}%
	}with $\mathrm{I_{n}}(k,a,N)$ defined as:{\normalsize \
	\begin{equation}
	I_{n}\left( k,a,N\right) =\left\{ {%
		\begin{array}{ccccccccccccccccccc}
		{1,} & {{\mathop{\rm Im}\nolimits}\left[ {{E_{n}}\left( k,a,N\right) }\right]
			>\varepsilon }  \\
		{0,} & {{\mathop{\rm Im}\nolimits}\left[ {{E_{n}}\left( k,a,N\right) }\right]
			\leq \varepsilon }
		\end{array}%
	}\right. ,
	\end{equation}%
	}where $\varepsilon $ is an arbitrarily chosen small number (we fix $%
	\varepsilon =10^{-6}$).
	
\begin{figure}[ht]
		{\normalsize \centering
			\includegraphics[width=0.5\linewidth,trim={0.4cm 1.8cm 1cm
				0.5cm},clip]{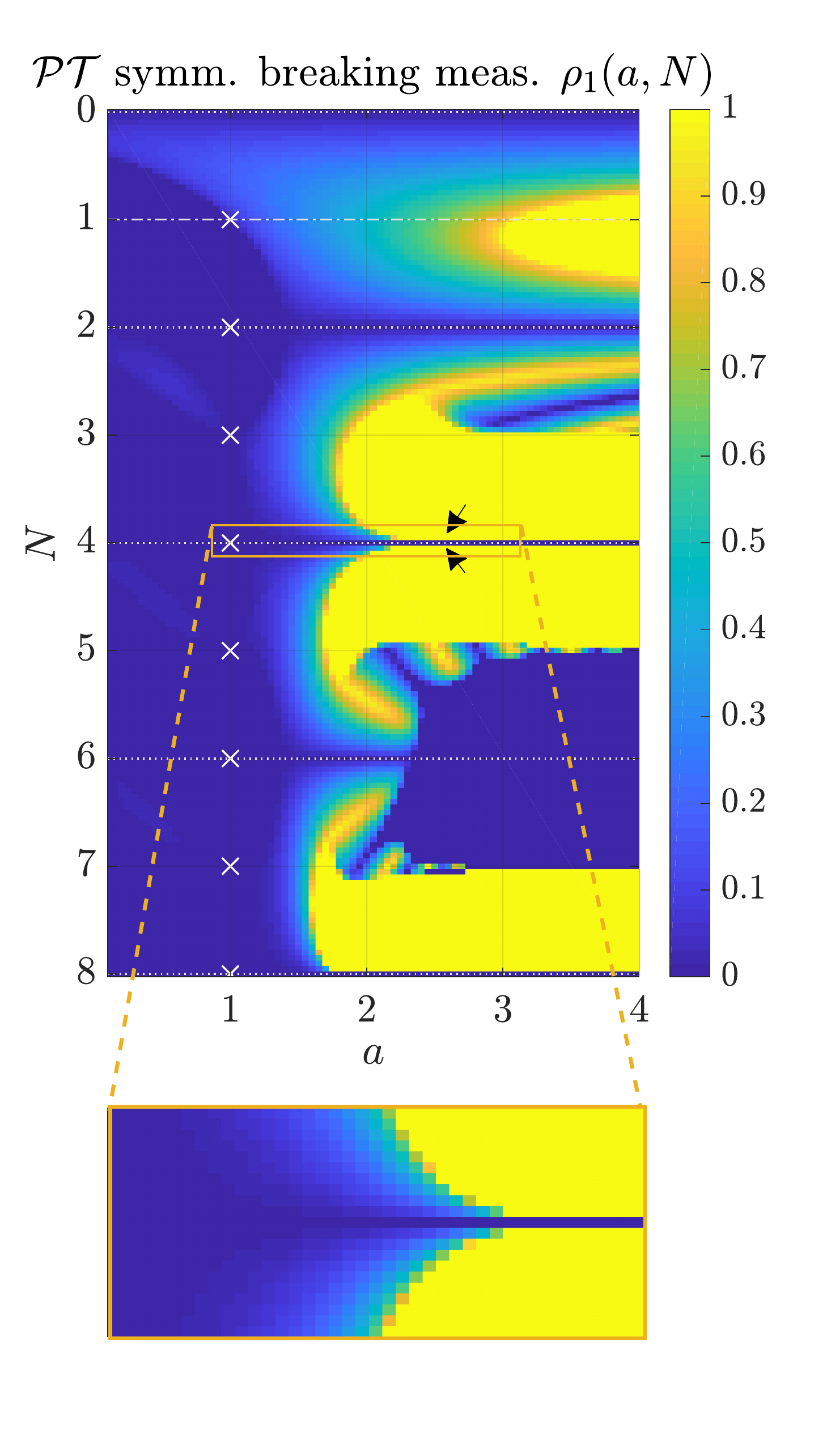} }
		\caption{ \textbf{The $\mathcal{PT}$ symmetry breaking measure $\mathbf{\
					\protect\rho _{1}(a,N)}$.} The darkest (brightest) color, corresponding to $%
			\protect\rho _{1}(a,N)=0$ ($\protect\rho _{1}(a,N)=1$), shows the domain of
			parameters where the first band is entirely real (complex). Some specific
			cases of interest are marked, including positive integer $N$ at $a=1$, where
			potential (\protect\ref{vsof}) is $\exp (iN\protect\xi )$ \textbf{(white
				crosses)}, even integer $N$ \textbf{(white dotted lines)}, $N=1$ \textbf{%
				(the white dashed dotted line),} and, finally, $N=4,\;a=2.5$ \textbf{(black
				arrows)}, where a sharp $\mathcal{PT}$ symmetry transition is clearly
			observed in the lower subframe zooming this region. 
		}
		\label{fig:measure}
	\end{figure}

	The measure was calculated in the first energy band for a set of parameter
	values in the range of $0<a\leq 4,\;0<N\leq 8$. This procedure produces a
	map indicating the relative degree of $\mathcal{PT}$-symmetry breaking within the
	examined range, which is displayed in Fig. \ref{fig:measure}. It shows that
	the spectrum remains real for even integer values of $N$ in Eq. (\ref{vsof}%
	), while it is obvious that symmetry breaking takes place when $N$ is an
	odd integer. Generally, as the parameter $a$ increases (starting at zero), a
	threshold is crossed at some point at which symmetry breaking sets in,
	while the further growth of this parameter increases the symmetry-breaking
	degree, until eventually all the real eigenvalues in the first band are
	eliminated. 
	
	Here, we focus on the analysis for the first (lowest) energy band, as $\mathcal{PT}$
	symmetry is, generally, more fragile in higher ones, making the situation
	less physically relevant. Nevertheless, some results for the second band are
	displayed below in Fig. \ref{fig:bands}.
	
	The particular case of potential (\ref{vsof}) with $a=1$ and integer $N$,
	i.e., $V_{\mathrm{SOF}}(\xi )=\exp (iN\xi )$, was studied previously \cite%
	{cannata1998schrodinger}. Further, for $N=1$ the latter potential is
	tantamount to the well-known one, $V=4\left[ \cos ^{2}(\xi )+ia\sin (2\xi )%
	\right] $, which has been examined in detail \cite%
	{PhysRevA.81.063807,makris2011mathcal} (the DC [constant] component in $V$
	may be eliminated by an overall energy shift). The map displayed in Fig. \ref%
	{fig:measure} demonstrates that, in some intervals of values of $N$, such as
	$0<N<2$, the increase in $a$ leads to crossing of the $\mathcal{PT}$%
	-symmetry-breaking threshold, while the breaking measure, $\rho _{1}$, keeps
	growing with the further increase of $a$. However, at other values of $N$
	the dependence on $a$ may be non-monotonous. For instance, at $N=2.5$ the
	further increase of $a$ beyond the symmetry-breaking threshold brings the
	system back to unbroken $\mathcal{PT}$-symmetry phase, with an entirely
	real energy spectrum. In this case, the subsequent increase of $a$ up to, at
	least, $a=8$ (the larges value for which the computation was performed) does
	not lead to a new symmetry-breaking event. In this connection, it is
	relevant to mention that examples of the restoration of the once broken $%
	\mathcal{PT}$-symmetry with the continuing increase of a relevant control
	parameter (typically, it is the strength of the gain-loss terms) are known
	in some completely different systems, such as nonlinear $\mathcal{PT}$%
	-symmetric models \cite{lumer2013nonlinearly}, nano-optical
	(subwavelength) $\mathcal{PT}$-symmetric media \cite{Huang14}, and in scattering and lasing models as well \cite{chong2011p,pilozzi2017topological}. Further,
	recall that even integer values of $N$ are exceptional, as the symmetry
	breaking does not take place for them.
	
	
	Our next observation is a very sharp transition between the $\mathcal{PT}$%
	-symmetric and broken-symmetry phases at $a>2.2$ in a vicinity of $N=4$. For
	$N=4$, the first energy band remains entirely real (as for all even $N$),
	while adding a small fractional part to $N$ turns the real band into a
	completely complex-valued one, as can be seen in Fig. \ref{fig:bands}, which
	displays the calculated energy eigenvalues in the first and second bands for
	potential (\ref{vsof}) with $a=2.5$ and $N=3.9$, $4.0$, and $4.1$. Note that
	the potentials seem almost identical in these three cases, while the
	difference in the energy-band structure is dramatic, including the $\mathcal{%
		PT}$-symmetry breaking taking place at $N=3.9$ and $4.1$, but absent at $N=4$%
	. The smallness of $\left\vert N-4\right\vert $ may be a reason for
	virtually constant values of the imaginary eigenvalues across the Brillouin
	zone, which may be a subject for additional analysis.{\normalsize \ }
	
	\begin{figure}[ht]
		{\normalsize \centering
			\includegraphics[width=0.7\linewidth,trim={0.6cm 1.3cm 0.5cm
				1.0cm},clip]{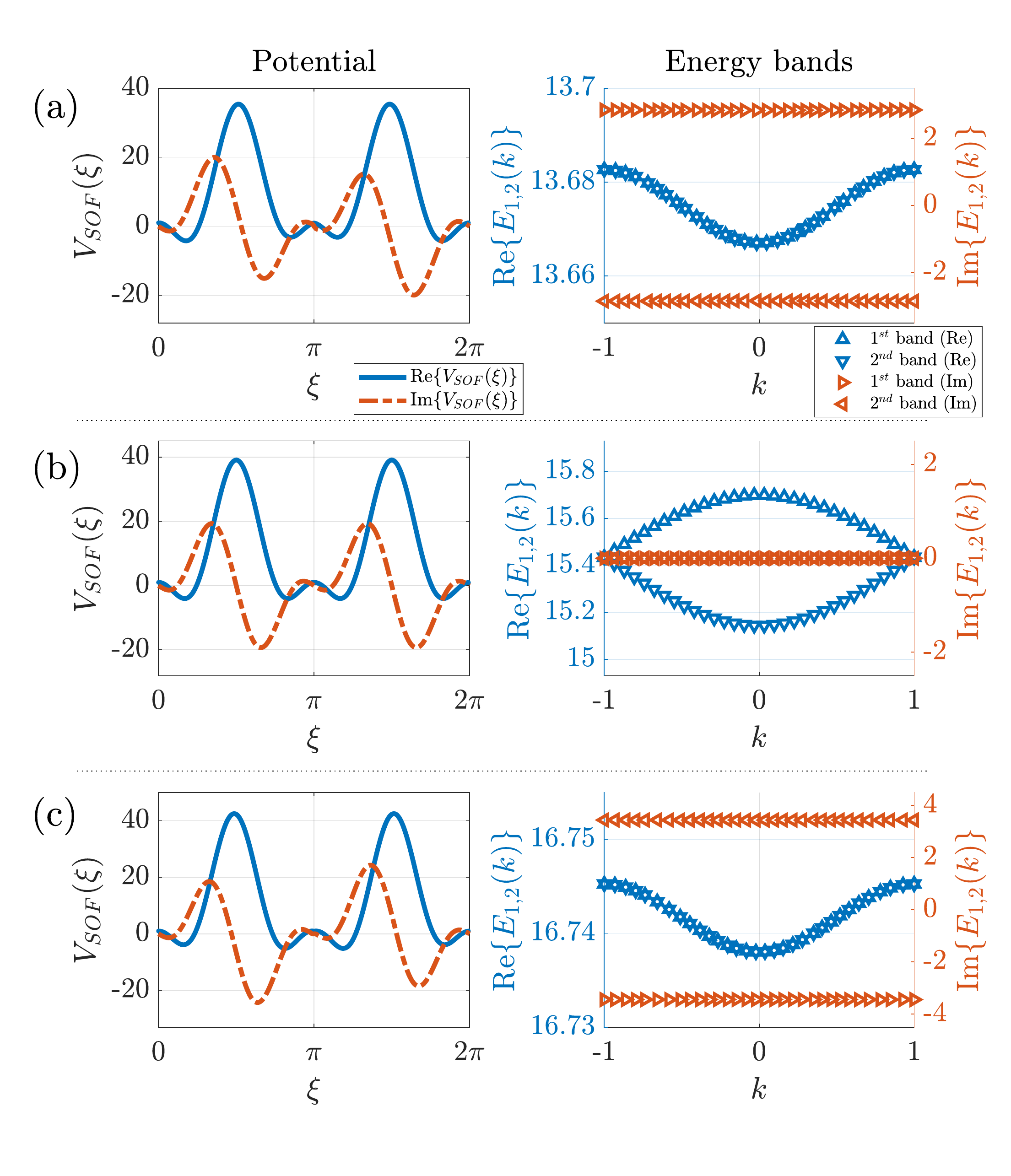}
		}
		\caption{ \textbf{The eigenvalue-band structure for selected potentials}.
			Here, $a=2.5$ while \textbf{(a)} $N=3.9$, \textbf{(b)} $N=4.0$, and \textbf{%
				(c)} $N=4.1$. \textbf{Left column:} The real (blue line) and imaginary (red
			dashed line) parts of each potential. \textbf{Right column:} The band
			diagrams showing the real and imaginary parts of the eigenvalues in the
			first and second bands. }
		\label{fig:bands}
	\end{figure}
	
	To further examine the sharp transition, we use a well-known technique \cite%
	{chicone2006,bender1999complex} to perform Floquet analysis of solutions
	to the Schr\"{o}dinger equation with potential (\ref{vsof}). Accordingly,
	stationary wave functions $\psi _{k}(\xi )$ are represented by a linear
	combination of two mutually orthogonal basis functions: $\psi _{k}(\xi
	)=c_{k}u_{1}(\xi )+d_{k}u_{2}(\xi )$, which are subject to the following
	boundary conditions:
	\begin{equation}
	\begin{array}{ccccccccccccccccccc}
	{{u_{1}}\left( 0\right) =1,} & {{u_{1}}^{\prime }\left( 0\right) =0,} \\
	{{u_{2}}\left( 0\right) =0,} & {{u_{2}}^{\prime }\left( 0\right) =1,}
	\end{array}
	\label{iconditions}
	\end{equation}%
	We have found the basis functions as numerical solutions to the stationary
	Schr\"{o}dinger equation, including potential (\ref{vsof}) with $a=2.5$ and $%
	N=3.9$, $4.0$,$~$or $4.1$. According to the Floquet analysis, the solution $%
	\psi _{k}(\xi )$, satisfying the Bloch condition, $\psi _{k}(\xi +\Lambda
	)=e^{ik\Lambda }\psi _{k}(\xi )$, with the potential's period $\Lambda $ (in
	the present notation, $\Lambda =\pi $), is bounded provided that the
	discriminant, $\Delta \left( E\right) \equiv u_{1}\left( \Lambda \right)
	+u_{2}^{\prime }\left( \Lambda \right) $, is real and meets the constraint $%
	\left\vert {\Delta \left( E\right) }\right\vert \leq 2$. When this criterion
	is satisfied, there exists a real-valued band of eigenvalues. Fig. \ref%
	{fig:floquet} shows the calculated discriminant $\Delta (E)$ as a function
	of the energy for each one of the three cases, $N=3.9$, $4.0$, and $4.1$. It is
	seen that, in the cases of $N=3.9$ and $N=4.1$ (the top and bottom rows) the
	discriminant's condition breaks for $E>0$, on the contrary to the case of $%
	N=4.0$, where a sharp minimum appears in the region of $6<E<7$, allowing for
	the boundedness of $\psi (\xi )$ and for the existence of a real-energy band.
	Further calculation of the real-energy bandwidth ($\max(E_n)-\min(E_n)$) for $N=4.0$ exhibits an exponential decrease as
	the $a$ parameter increases. We attribute the narrowing of the band to the exponential decrease in the magnitude of the superoscillatory feature in the imaginary part of the potential, and to the increase of the maxima magnitude in the real part.
	
	\begin{figure}[h]
		{\normalsize \centering
			\includegraphics[width=0.65\linewidth,trim={0.2cm 2.0cm 1.8cm
				1.0cm},clip]{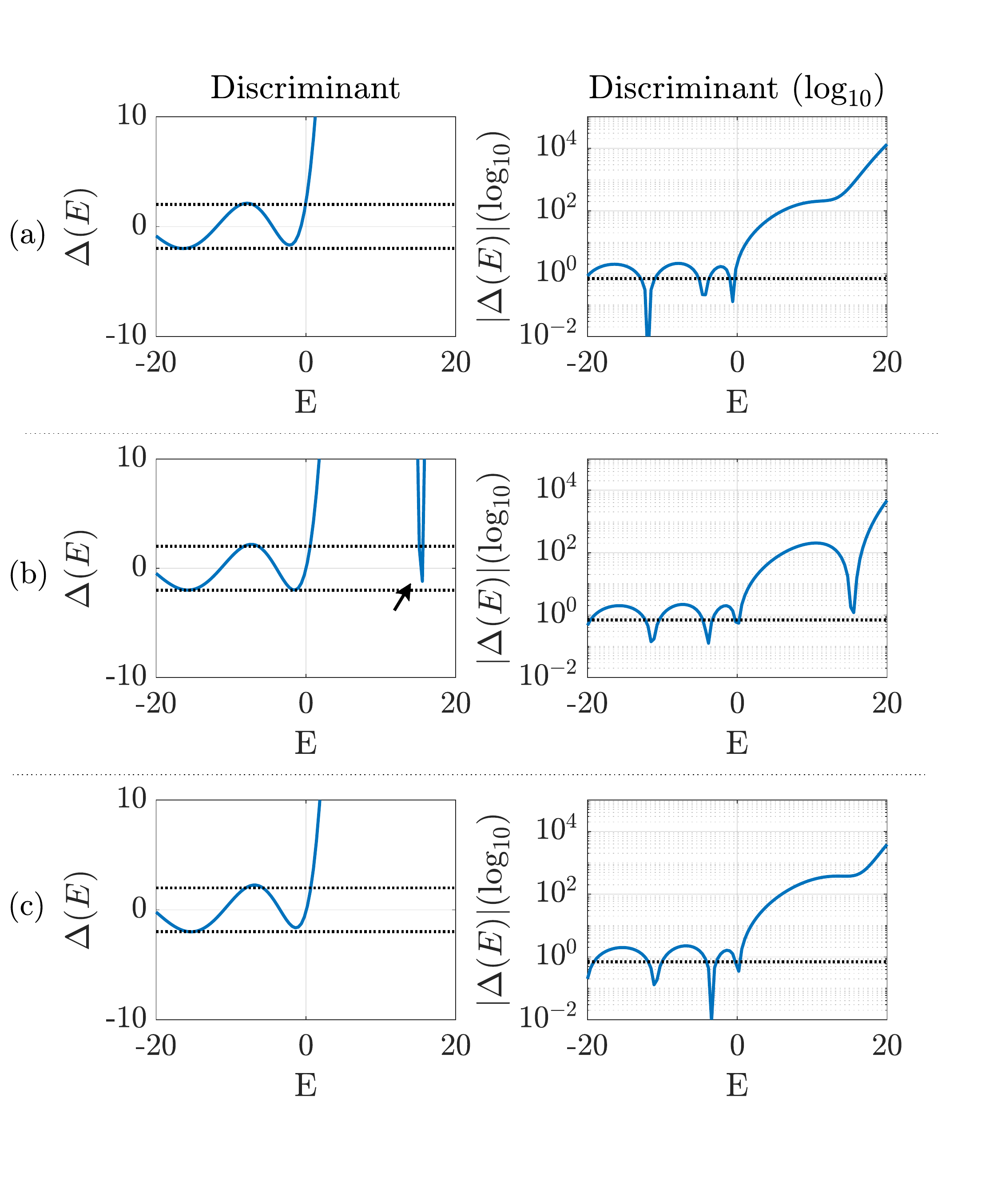} }
		\caption{ \textbf{The calculated discriminant function $\mathbf{\Delta (E)}$}
			, which determines the existence of Bloch wave functions corresponding to
			real eigenvalues, for $a=2.5$ and \textbf{(a)} $N=3.9$, \textbf{(b)} $N=4.0$%
			, \textbf{(c)} $N=4.1$. Dotted vertical lines mark the boundaries $\Delta
			(E)=\pm 2$ indicating the existence area for the wave functions. The arrow
			marks a sharp minimum at $E\approx 15.56$. }
		\label{fig:floquet}
	\end{figure}
	
\begin{figure}[tbp]
		{\normalsize \centering
			\includegraphics[width=0.9\linewidth,trim={4.0cm 0.9cm 4.2cm
				1.2cm},clip]{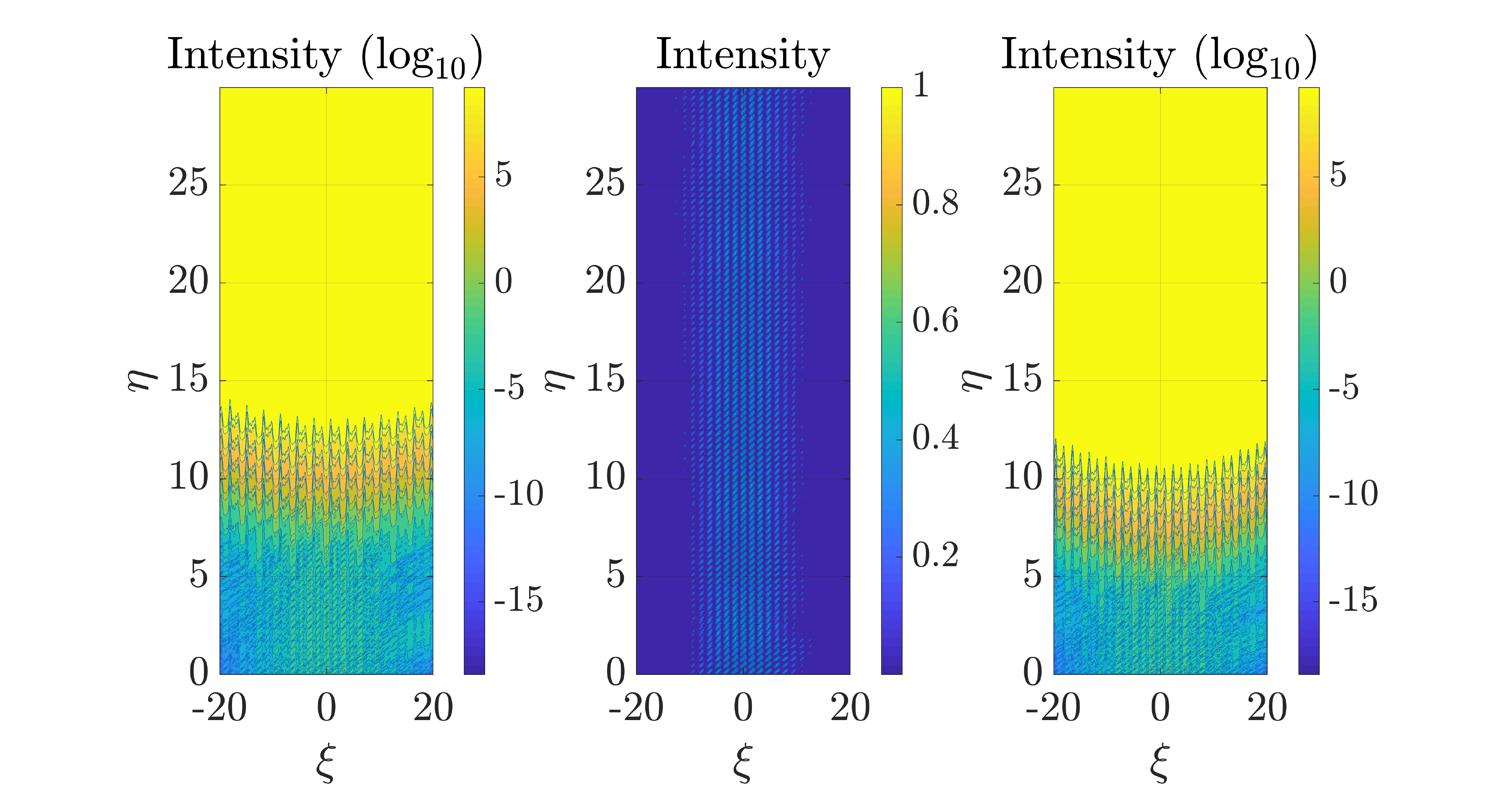}
		}
		\caption{\textbf{The simulated evolution of the local intensity of the broad
				initial Gaussian, displayed for potential (\protect\ref{vsof}) with }$a=2.5$
			\textbf{\ and }$N=3.9$\textbf{\ (left), }$N=4.0$\textbf{\ (center), }$N=4.1$
			\textbf{\ (right). }In the left and right panels, the log scale is used.}
		\label{fig:evolution}
	\end{figure}

	Next, we used the Crank-Nicolson algorithm \cite{Crank1996} to simulate the
	evolution of a wide Gaussian wave packet, set initially around $\eta =0$,
	governed by Eq. (\ref{schrodinger}), again with the potential (\ref{vsof})
	corresponding to $a=2.5$ with $N=3.9$, $4.0$, and $4.1$. Figure \ref%
	{fig:evolution} displays the evolution of the local intensity of the wave
	packet, $\left\vert \psi \left( \xi ,\eta \right) \right\vert ^{2}$, in each
	case. The blowup (exponential growth of the field's amplitude) is, quite
	naturally, observed in the cases of broken {\normalsize $\mathcal{PT}$ }%
	symmetry, corresponding to $N=3.9$ and $4.1$ (the left and right panels in
	Fig. \ref{fig:evolution}), while, in the absence of symmetry breaking ($%
	N=4.0$, the central panel), the wave packet retains a stable shape.
	
	To complete the picture presented in Fig. \ref{fig:measure}, we finally
	consider the case of $a=0$ and $0\leq N\leq 8$, i.e., the potentials in the
	form of $V_{\mathrm{SOF}}(\xi )=\cos ^{N}(2\xi )$. In the case when $N$ is
	an integer, this potential is clearly Hermitian, generating real energy
	spectra, while when $N$ is a rational fraction, the potential is multivalued
	and generally complex. Yet, unlike the case of $a\neq 0$, where the entire
	potential is complex-valued, a fractional power of $\cos (2\xi )$ produces
	roots which are entirely real or piecewise real and complex. For any
	positive value of $\cos (2\xi )$, real roots always exist, while for $\cos
	(2\xi )<0$ a real root exists if $N$ is represented by an irreducible
	rational fraction, whose denominator is an odd integer:%
	\begin{equation}
	N=P/Q,\quad ~Q=1+2M,  \label{PQ}
	\end{equation}%
	($M$ is an arbitrary integer).
	Thus, one can construct the potential as a set of real roots, in case they
	are available, adding complex roots with the smallest phase when real roots
	are absent. Naturally, when Eq. (\ref{PQ}) holds, the entire potential is real, producing a fully real eigenvalue spectrum. On the other hand, when $Q$ in expression (\ref{PQ}) is even, the potential includes complex
	segments, which results in $\mathcal{PT}$ symmetry breaking
	in the entire first band. The conclusion is that condition (\ref{PQ}), the validity of which has a sparse structure with respect to the
	rationals and is discontinuous everywhere, determines $\rho _{1}(0,N)$ as
	follows:
	\begin{equation}
	\rho _{1}\left( {0,N}\right) =\left\{ {%
		\begin{array}{ccccccccccccccccccc}
		{0,} & N=P/(1+2M),  \\
		{1,} & \mathrm{otherwise},
		\end{array}%
	}\right.   \label{rhoazero}
	\end{equation}%
	\\
	which implies that the $\mathcal{PT}$ symmetry breaking measure $\rho
	_{1}(0,N)$ itself is discontinuous and sparse. \\
	
	\section{\label{sec:level2} Conclusions}
	
	We have examined the properties of the canonical suboscillatory and
	superoscillatory complex wave functions from quantum mechanics, which are
	given by Eqs. (\ref{eq2}) and (\ref{vsof}), respectively, as {\normalsize $%
		\mathcal{PT}$}-symmetric potentials in the one-dimensional Schr\"{o}dinger
	equation. In the former case, we have found that such a complex potential
	always generates a purely real spectrum of energy eigenvalues, avoiding
	$\mathcal{PT}$ symmetry breaking. A more interesting
	situation takes place in the latter case, where the complex potential (\ref%
	{vsof}) gives rise to intricate phenomenology of the {\normalsize $\mathcal{%
			PT}$}-symmetry breaking. Depending on values of the parameters, $a$ and $N$
	in Eq. (\ref{vsof}), we have found a wide region in which the symmetry
	breaking develops smoothly, with the increase of the control parameter $a$,
	and, on the other hand, a region exhibiting an extremely sharp transition
	from the phase of unbroken{\normalsize \ $\mathcal{PT}$} symmetry to broken
	symmetry. Another noteworthy finding is a possibility of the restoration of
	the originally broken {\normalsize $\mathcal{PT}$} symmetry with subsequent
	growth of the control parameter $a$. Direct simulations of the evolution of
	input field pulses demonstrate their stability in the case of the unbroken
	symmetry, and a blowup when the symmetry was broken. Generally, our analysis
	shows that two families of complex potentials offer an essential extension
	of previously examined {\normalsize $\mathcal{PT}$}-symmetric ones, and
	suggests that the new potentials may find application to waveguiding,
	lasing, filtering, and optical sensing. The refractive-index and gain-loss
	profiles emulating these potentials can be created by means of available
	experimental techniques.
	
	As an extension of the present work reported, it may be interesting to consider
	a model combining the new {\normalsize $\mathcal{PT}$}-symmetric potentials
	with nonlinearity of the optical medium. A challenging direction for the
	development of the present analysis may be its extension to
	two-dimensional geometry.

	\nolinenumbers
	
	\bibliographystyle{ieeetr}
	\bibliography{bibliography}
	
\end{document}